\documentclass[a4paper,english,aps,preprint,superscriptaddress,eqsecnum]{revtex4}

\usepackage{graphicx,amsfonts,amsmath}
\usepackage{hyperref}

\renewcommand{\a}{\alpha}
\renewcommand{\b}{\beta}
\newcommand{\g}{\gamma}

\def\m{\mu}

\def\q{\theta}
\def\r{\rho}

\def\L{\Lambda}

\begin{document}

\title{On the duality of three-dimensional superfield theories}

\author{A. F. Ferrari}
\author{M. Gomes}
\affiliation{Instituto de F\'\i sica, Universidade de S\~ao Paulo\\
Caixa Postal 66318, 05315-970, S\~ao Paulo, SP, Brazil}
\email{alysson,mgomes,ajsilva@fma.if.usp.br}
\author{J. R. S. Nascimento}
\author{A. Yu. Petrov}
\affiliation{Departamento de F\'{\i}sica, Universidade Federal da Para\'{\i}ba\\
 Caixa Postal 5008, 58051-970, Jo\~ao Pessoa, Para\'{\i}ba, Brazil}
\email{jroberto,petrov@fisica.ufpb.br}
\author{A. J. da Silva}
\affiliation{Instituto de F\'\i sica, Universidade de S\~ao Paulo\\
Caixa Postal 66318, 05315-970, S\~ao Paulo, SP, Brazil}
\email{alysson,mgomes,ajsilva@fma.if.usp.br}

\begin{abstract}
Within the superfield approach, we consider the duality between the supersymmetric Maxwell-Chern-Simons and self-dual theories in three spacetime dimensions. Using a gauge embedding method, we construct the dual theory to the self-dual model interacting with a matter superfield, which turns out to be not the Maxwell-Chern-Simons theory coupled to matter, but a more complicated model, with a ``restricted'' gauge invariance. We stress the difficulties in dualizing the self-dual field coupled to matter into a theory with complete gauge invariance. After that, we show that the duality, achieved between these two models at the tree level, also holds up to the lowest order quantum corrections.  \end{abstract}

\maketitle

\section{Introduction}
\label{intro}

For a long time, it has been recognized  that it is  important to establish connections between apparently unrelated situations so that unifying pictures may emerge. In this context, the duality between the Abelian Maxwell-Chern-Simons (MCS) and self-dual (SD) theories in three dimensional spacetime found in~\cite{nieu} is a paradigmatic example. Extensions of this relation involving non-Abelian gauge fields were considered by various authors~\cite{dualym}. However, while the duality is well established for the free case, the situation becomes more subtle when interactions with others dynamical fields are taken into account. For instance, in~\cite{malacarne} a master field action was used to show that the duality between the MCS and SD models coupled to fermions requires the addition of a Thirring current-current interaction in the MCS Lagrangian. However, problems were met in using the same method to study the duality when interactions with a bosonic field were present. In~\cite{ilha,anacleto}, a so-called gauge embedding procedure was developed to overcome these problems. In this way, it became possible to build a theory dual to the SD model coupled to bosons or fermions. In the later case, this dual model turned out to be the MCS theory with a Thirring interaction, as found in~\cite{malacarne}, while in the former case a more complicated situation arose. Indeed, the theory dual to the SD model coupled to bosons was found to be a modified MCS theory with an unusual field-dependent coefficient for the Maxwell term.

Another interesting question concerns the realization of the duality for supersymmetric models. A first step in this direction was the use of the master field action approach in~\cite{nieu2} to study the equivalence between the supersymmetric MCS and SD theories in the superfield formulation. However, using this method, it was not possible to go beyond the simplest case of Abelian theories without any coupling to matter. Our aim here is to propose a generalization of the gauge embedding method to construct a theory dual to the supersymmetric self-dual model interacting with a scalar superfield. We will show that this dual theory involves both a Thirring interaction, as well as the modified MCS part. After building the duality at the classical level, we will show that it survives when the first order quantum corrections are taken into account. 

It is important to remind that this modified MCS Lagrangian does not define a genuine gauge theory in the usual sense, since its action is invariant under a ``restricted'' gauge invariance, that is to say, when only the basic spinor superpotential undergoes a gauge transformation. As we shall see, it is very difficult to generalize the gauge embedding method in order to turn the dual of the SD model into a genuine gauge theory.

This paper is organized as follows. In Section~\ref{freecase}, we review the duality between supersymmetric SD and MCS models, without coupling to dynamical sources, using the superfield formalism. Afterwards, in Section~\ref{gaugeembedding}, we include the interaction of the SD with a matter superfield, and use the gauge embedding method to build the dual of this theory. The dual equivalence so obtained is shown to be maintained by the lowest-order quantum corrections in Section~\ref{quantumduality}. In Section~\ref{completege}, we describe the difficulties that arise when one tries to find a dual for the SD model which is a ``genuine'' gauge theory. Our conclusions, together with some comments on the applicability of these methods to the noncommutative extensions of these models, are found in Section~\ref{conclusions}.

\section{Duality for the free theories}
\label{freecase}

Our starting point is the superfield Maxwell-Chern-Simons theory which is described by the action 
\begin{eqnarray}
\label{mcs}
S_{MCS}\,=\,-\int d^5 z(W^{\a}W_{\a}-2mW^{\a}A_{\a}),\quad\a=1,2\,,
\end{eqnarray}

\noindent
where
\begin{eqnarray}
\label{w}
W^{\a}\,=\,\frac{1}{2} D^{\b}D^{\a}A_{\b}
\end{eqnarray}

\noindent
is the usual superfield strength and $A_\a$ is the spinor superpotential. Hereafter, we follow the conventions of~\cite{SGRS}. The action~(\ref{mcs}) is invariant under the gauge transformation
$\delta A_{\a}=D_{\a}\epsilon$. After the addition of the gauge-fixing term
\begin{eqnarray}
S_{gf}=-\frac{1}{2\xi}\int d^5 z D^{\b}A_{\b}D^2D^{\a}A_{\a}\,,
\end{eqnarray}

\noindent
the propagator of the $A^{\a}$ field takes the form
\begin{eqnarray}
\label{pa}
<A^{\a}(-p,\q_1)A^{\b}(p,\q_2)> \,= \, -\frac{i}{4}
\left[\frac{(D^2+2m)D^{\b}D^{\a}}{p^2(p^2+4m^2)}-\xi\frac{D^2D^{\a}D^{\b}}{p^4}
\right]\delta_{12},
\end{eqnarray}

\noindent
were $\delta_{12}\equiv\delta^2(\theta_1 - \theta_2)$. From Eq.~(\ref{pa}), we obtain the propagator of the $W^{\a}$ superfield,
\begin{eqnarray}
\label{pw}
<W^{\a}(-p,\q_1)W^{\b}(p,\q_2)>&=&\frac{1}{4}
\overset{\longrightarrow}{D^{\gamma}D^{\a}}
<A_{\gamma}(-p,\q_1)A_{\delta}(p,\q_2)>
\overset{\longleftarrow}{D^{\b}D^{\delta}}\nonumber\\
&=& \frac{i}{4}\frac{(D^2+2m)D^{\beta}D^{\alpha}}{p^2+4m^2}\delta_{12}\,.
\end{eqnarray}

Let us now consider the self-dual theory whose action is specified by
\begin{eqnarray}
\label{sd}
S_{SD}\,=\,4\int d^5 z \, \left(m^2B^{\a}B_{\a}- \frac{1}{2} m \Omega^{\a}B_{\a}\right),
\end{eqnarray}

\noindent
where $B^{\a}$ is a spinor superfield and $\Omega^{\a}=\frac{1}{2} D^{\b}D^{\a}B_{\b}$ is an analog of the superfield strength defined in the MCS theory. The propagator of the $B^{\a}$ superfield is
\begin{equation}
\label{pb}
<B^{\a}(-p,\q_1)B^{\b}(p,\q_2)>\,=\,i
\Big[\frac{D^2(D^2+2m)D^{\b}D^{\a}}{8mp^2(p^2+4m^2)}-
\frac{D^2D^{\a}D^{\b}}{16m^2p^2}
\Big]\delta_{12}.
\end{equation}

\noindent
Notice that the propagators for $W^{\a}$ and $B^{\a}$ both have a pole at $p^2=-4m^2$. Near this pole the $B^{\a}$ propagator becomes
\begin{eqnarray}
\label{pbfin}
<B^{\a}(-p,\q_1)B^{\b}(p,\q_2)>&=&
\frac{i}{16}\frac{(D^2+2m)D^{\beta}D^{\alpha}}{m^2(p^2+4m^2)}\delta_{12}
+\cdots,
\end{eqnarray}

\noindent
where the dots stand for terms which stay finite as $p^2\to -4m^2$.
Thus the superfield $B^{\a}$ of the supersymmetric self-dual model
seems to play the same role as the superfield strength $\frac1{2m}W^{\a}$ of the
MCS theory. This conclusion is further substantiated by the equations of motion derived from the actions in~(\ref{mcs}) and~(\ref{sd}),
\begin{subequations}
\begin{eqnarray}
& &-mD^{\b}D^{\a}B_{\b}+4m^2B^{\a}\,=\,0 \,, \\
& &-D^{\b}D^{\a}W_{\b}+4mW^{\a}\,=\,0 \,.
\end{eqnarray}
\end{subequations}

Denoting the vector components of the superfields $W^{\a}$ and $B^{\a}$ respectively as $f_m = \frac{1}{2}(\gamma_m)^{\a \b} f_{\a\b}$ and $V_m = \frac{1}{2}(\gamma_m)^{\a \b} V_{\a\b}$, where $m=0,1,2$ and
\begin{subequations}
\begin{eqnarray}
V_{\a\b}&=&-\frac{i}{2}D_{(\a}B_{\b)}|_{\theta=0} \, ,\\
f_{\a\b}&=&\frac{1}{2}D_{(\a}W_{\b)}|_{\theta=0} \, ,
\end{eqnarray}
\end{subequations}

\noindent
we find that the propagators for these fields in the neighborhood of the above mentioned pole coincide,
\begin{equation}
\label{pcom}
<V^m(-p)V^n(p)>\,=\,\frac{1}{4m^2}<f^m(-p)f^n(p)>\,=\,
\frac{i}{2}\frac{(p^mp^n-\eta^{mn}p^2-2m\epsilon^{mnl}p_l)}{4m^2(p^2+4m^2)}\,.
\end{equation}

\noindent
We remind the reader that $f^m$ is the dual of the $F_{mn}$ tensor, the field strength of the ``electromagnetic'' component field in $A_\a$.

\section{Interaction with a scalar superfield and the gauge embedding method}
\label{gaugeembedding}

Here we investigate the persistence of the duality pointed out in the previous section 
when interaction with matter is included. In this situation the approach developed in~\cite{malacarne}, based on the use of the Green functions for the field equations, cannot be directly applied. In fact, the coupling of the gauge superfield to the (scalar) matter superfield cannot be represented in the form $A^{\a}J_{\a}$ because of the presence of an extra ``diamagnetic'' term characteristic of the minimal coupling. To circumvent this problem we use a gauge embedding approach, similar to the one developed in~\cite{anacleto}. 

Let us introduce the action of the self-dual model coupled to matter,
\begin{eqnarray}
\label{sdm}
S_{SDM}&=&\int d^5 z\Big\{ 4m^2B^{\a}B_{\a}-m
B^{\a}D^{\b}D_{\a}B_{\b}-\nonumber\\
&-&
\frac{1}{2}\left[
\left(D^{\a}\bar{\phi}+igB^{\a}\bar{\phi}\right)
\left(D_{\a}\phi-ig\phi B_{\a}\right)
+2M\phi\bar{\phi}\right]
\Big\}.
\end{eqnarray}

\noindent
The pure matter sector of this theory is invariant under the transformations $\phi\to\phi e^{i\epsilon}$,
$\bar{\phi}\to e^{-i\epsilon}\bar{\phi}$, $B_{\a}\to
B_{\a}+\frac{1}{g}D_{\a}\epsilon$. 

Our aim consists in transforming the whole self-dual theory (\ref{sdm}) into a gauge theory, in a sense to be clarified later. We start by recasting the pure matter sector of the action~(\ref{sdm}) in the form,
\begin{eqnarray}
S_m\,=\,\int d^5 z \left[\phi(D^2-M)\bar{\phi}+
B^{\a}J_{\a}-
\frac{g^2}{2}\bar{\phi}B^{\a}B_{\a}{\phi} \right],
\end{eqnarray}

\noindent
where we used the notation
\begin{eqnarray}
\label{current}
J_{\a}=i\frac{g}{2}\phi\stackrel{\leftrightarrow}{D}_{\a}\bar{\phi}\,.
\end{eqnarray}

\noindent
The ``gauge'' current of the model differs from the above expression by a $B$ field dependent term,
\begin{eqnarray}
{\cal J}_{\a}&\equiv&\frac{\delta S_m}{\delta B_{\a}}=
J_{\a}-\bar{\phi}B_{\a}\phi \, \equiv \,
i\frac{g}{2}\phi\stackrel{\leftrightarrow}{\nabla}_{\a}\bar{\phi}\,,
\end{eqnarray}

\noindent
and the equations of motion for the gauge and scalar superfields are
\begin{subequations}
\begin{eqnarray}
& &8\m^2B_{\a}-2mD^{\b}D_{\a}B_{\b}+J_{\a}\,=\,0\label{emsd1}\,,\\
& & (D^2-M)\phi-igB^{\a}D_{\a}\phi-\frac{g^2}{2}B^{\a}B_{\a}\phi\,=\,0\label{emsd2}\,.
\end{eqnarray}
\end{subequations}

\noindent
Here $\m^2=m^2-\frac{g^2}{8}\phi\bar{\phi}$ is the field-dependent ``mass''
of the $B^{\a}$ superfield. We will refer to $\frac{\delta S}{\delta B^{\a}}\equiv K_{\a}$, given by the left-hand side of~(\ref{emsd1}), as the Euler vector
\begin{eqnarray}
\label{euler}
K_{\a}\,=\,8\m^2B_{\a}-2mD^{\b}D_{\a}B_{\b}+J_{\a}\,,
\end{eqnarray}

\noindent
and we note that $K_{\a}$ can be rewritten in terms of the ``gauge'' current ${\cal J}_{\a}$ as
\begin{eqnarray}
K_{\a}\,=\,m^2B_{\a}-2mD^{\b}D_{\a}B_{\b}+{\cal J}_{\a}\,.
\end{eqnarray}

The gauge embedding procedure is an iterative method that starts by the introduction of an auxiliary field $\L_{\a}$, which is a Lagrange multiplier for the Euler vector corresponding to the spinor superfield (the introduction of the iterative method with respect to both the spinor and scalar superfields, which in principle would provide complete gauge invariance in the resulting model, becomes much more complicated, as we show in Section~\ref{completege}). We therefore define the first-order iterated Lagrangian
\begin{eqnarray}
L^{(1)}\,=\,L_{SDM}-\L^{\a}K_{\a}\,.
\end{eqnarray}

\noindent
The variation of $L^{(1)}$ under the transformations $B_{\a} \to B_{\a}+D_{\a}\epsilon$
gives
\begin{eqnarray}
\delta L^{(1)}\,=\,K^{\b}D_{\b}\epsilon-K^{\a}\delta\L_{\a}-\L^{\a}\delta K_{\a}\,,
\end{eqnarray}

\noindent
while the change in the Euler vector $K_\a$ is
\begin{equation}
\label{compare}
\delta K_\a \, = \, 8 \m^2 D_\a \epsilon\,,
\end{equation}

\noindent
and, therefore, if we define $\delta\L_{\a}=D_{\a}\epsilon$, the variation $\delta L^{(1)}$ turns out to be
\begin{eqnarray}
\delta L^{(1)}\,=\,-\L^{\a}\delta K_{\a}=-8\m^2\L^{\a}D_{\a}\epsilon\,=\,
-4\m^2\delta(\L^{\a}\L_{\a})\,,
\end{eqnarray}

\noindent
so that it can be canceled by the variation of the term $4\m^2\L^{\a}\L_{\a}$. Thus, the second-order iterated Lagrangian
\begin{eqnarray}
L^{(2)}\,=\,L_{SDM}-\L^{\a}K_{\a}+4\m^2\L^{\a}\L_{\a}
\end{eqnarray}

\noindent
is invariant under the gauge transformation $\delta B_{\a}=D_{\a}\epsilon$. The Lagrange multiplier $\L_{\a}$ can be eliminated using its equation of motion
\begin{eqnarray}\label{24}
K_{\a}\,=\,8\m^2\L_{\a}\,,
\end{eqnarray}

\noindent
and we finally arrive at the gauge invariant Lagrangian
\begin{eqnarray}
L^{(2)}\,=\,L_{SDM}-\frac{1}{16\m^2}K^{\a}K_{\a}\,,
\end{eqnarray}

\noindent
whose explicit form is
\begin{eqnarray}
L^{(2)}&=&\Big\{ 4m^2B^{\a}B_{\a}-m
B^{\a}D^{\b}D_{\a}B_{\b}-\nonumber\\&-&
\frac{1}{2}[(D^{\a}\phi-ig\phi B^{\a})(D_{\a}\bar{\phi}+igB_{\a}\bar{\phi})+
2M\phi\bar{\phi}]
\Big\}-\nonumber\\&-&
\frac{1}{16\m^2}\left(8\m^2B^{\a}-2mD^{\b}D^{\a}B_{\b}+J^{\a}\right)
\left(8\m^2B_{\a}-2mD^{\g}D_{\a}B_{\g}+J_{\a}\right)\,.
\end{eqnarray}

\noindent
After renaming the spinor superfield in the previous equation as $A_\a$ and some rearrangements, we can cast the action we have found for the theory dual to the SD Lagrangian in~(\ref{sdm}) as
\begin{eqnarray}
\label{new}
L_{DMCS}&=&\phi(D^2-M)\bar{\phi}+2mA^{\a}W_{\a}-\frac{m^2}{\m^2}W^{\a}W_{\a}
-\frac{1}{16\m^2}J^{\a}J_{\a}+\frac{m}{2\m^2}W^{\a}J_{\a}\,,
\end{eqnarray}

\noindent
where the superfield strength $W^{\a}$ has been defined in Eq.~(\ref{w}) and the $J_{\a}$ is given in~(\ref{current}). In the pure spinor sector, the Lagrangian~(\ref{new}) is similar to the superfield Maxwell-Chern-Simons action. However, the $W^\a W_\a$ term has an unconventional field dependent coefficient, as it happens in some generalizations of the  Abelian Higgs model (see for instance \cite{Kao}). We stress again that the action obtained from Eq.~(\ref{new}) is invariant under gauge transformations of the $A_\a$ superfield alone. Despite all this, we will refer to Eq.~(\ref{new}) as a dualized Maxwell-Chern-Simons (DMCS) model. Another point worth of noticing is the generation of a Thirring interaction. We also remark that the gauge field interaction with the matter given by the term $W^{\a}J_{\a}$ is the superfield analog of the ``magnetic'' coupling $\epsilon_{abc}\partial^a A^b J^c$~\cite{malacarne}.

Let us now compare the equations of motion for the spinor superfield in the self-dual model~(\ref{sdm}) and for the superfield strength in the DMCS model~(\ref{new}). After introducing the operator
\begin{equation}
\label{deltam1}
\left(\Delta^{-1}\right)_\a^{\,\,\b} \, \equiv \, 8(\m^2 \delta_\a^{\,\,\b} -\frac{1}{4}m D^{\b}D_{\a}) \, ,
\end{equation}

\noindent
they are given respectively by
\begin{subequations}
\begin{align}
&\left(\Delta^{-1}\right)_\a^{\,\,\b} \, B_\b \, = \, - J_\a \, , \label{eqofm1} \\
&\left(\Delta^{-1}\right)_\a^{\,\,\b} \, \left( \frac{W_\b}{\m^2}\right) \, = \, -\frac{1}{2}D^\b D_\a \left( \frac{J_\b}{\m^2} \right) \,. \label{eqofm2}
\end{align}
\end{subequations}

Given the inverse of the $\left(\Delta^{-1}\right)_\a^{\,\,\b}$ operator as
$\Delta_\r^{\,\,\a} \, \left(\Delta^{-1}\right)_\a^{\,\,\b}  =  \delta_\r^{\,\,\b}$, the solution of Eq.~(\ref{eqofm1}) can be readily obtained,
\begin{equation}
\label{solu1}
B_\a \, = \, -\Delta_\a^{\,\,\b} \, J_\b \,,
\end{equation}

\noindent
while, for solving Eq.~(\ref{eqofm2}), one starts by applying $\Delta_\r^{\,\,\a}$ to Eq.~(\ref{deltam1}) to obtain
\begin{equation}
\frac{1}{2} \, \Delta_\r^{\,\,\a} D^{\b}D_{\a} \, = \, \frac{2}{m} \Delta_\r^{\,\,\a}\,\m^2 \, - \frac{1}{4m} \delta_\r^{\,\,\a} \,,
\end{equation}

\noindent
which can be used to write the solution of Eq.~(\ref{eqofm2}) as
\begin{equation}
\label{solu2}
\frac{W_\a}{\m^2} \, = \, - \frac{2}{m} \Delta_\a^{\,\,\b}\,J_\b \, + \frac{1}{4 \m^2 m} J_\a \,.
\end{equation}

\noindent
By comparing Eqs.~(\ref{solu1}) and~(\ref{solu2}), we conclude that
\begin{eqnarray}
\label{expw}
W_\b\,=\, \frac{J_{\b}}{4 m} + \frac{2 \mu^2}{m} B_\b \,,
\end{eqnarray}

\noindent 
and therefore as $g\rightarrow 0$ one recovers the relation $B^{\alpha}= \frac{W^{\alpha}}{2m}$.

It still remains to verify the equivalence for the matter sectors of these models. To this end we consider the equation of motion for the scalar superfield $\phi$ corresponding to the DMCS model,
\begin{align}
(D^2-M)\phi +  &\frac{m^2}{\mu^4}W^{\a}W_{\a}\frac{\partial \mu^2}{\partial\overline{\phi}}+\frac{1}{16\mu^4}J^{\a}J_{\a}\frac{\partial \mu^2}{\partial\overline{\phi}}
-\frac{1}{8\mu^2}J^{\a}\frac{\partial J_{\a}}{\partial\overline{\phi}} \nonumber\\
&\frac{m}{2\mu^4}\frac{\partial \mu^2}{\partial\overline{\phi}}W^{\a}J_{\a}+
\frac{m}{2\mu^2}W^{\a}\frac{\partial J_{\a}}{\partial\overline{\phi}}=0.
\end{align}

\noindent
By using the expression~(\ref{expw}) we arrive at
\begin{eqnarray}
(D^2-M)\phi-igB^{\a}D_{\a}\phi-\frac{g^2}{2}B^{\a}B_{\a}\phi\,=\,0\,,
\end{eqnarray}

\noindent
which coincides with the equation of motion for the matter superfield in the SD model. This confirms the complete duality equivalence of these two models.

The Lagrangian in Eq.~(\ref{new}) contains nonrenormalizable
interactions but, concerning renormalizability, we do not expect these to generate difficulties at the quantum level. As it happens in the nonsupersymmetric model~\cite{gomes}, the Thirring interaction is renormalizable in the framework of the $\frac{1}{N}$ expansion 
for a $N$-component scalar superfield. Indeed, in that case we can eliminate the four-scalar vertex $J^{\a}J_{\a}$ in favor of $S^{\a}\phi_i\stackrel{\leftrightarrow}{D}_{\a}\bar{\phi}_i-
\frac{1}{2}S^{\a}S_{\a}$ where $S_{\a}$ is an auxiliary superfield, whose propagator is, up to a constant, equal to the one for the
gauge spinor superfield in the $CP^{N-1}$ model \cite{cpn}. It
behaves as $1/k$ for large $k$ momentum, drastically improving the
power counting. One may even entertain the hope that renormalizability also holds for finite $N$, although if a direct proof is not feasible at the moment.

\section{Inclusion of Radiative Corrections}
\label{quantumduality}

After establishing the duality between the SD model defined by Eq.~(\ref{sdm}) and the DMCS model in Eq.~(\ref{new}) at the classical level, we shall now present some calculations to verify whether this equivalence persists at the quantum level. In more concrete terms, we will examine the radiative corrections to the two point vertex functions of the $B_{\a}$, $W_{\a}$ and $\phi$ superfields for both theories, up to the second order in the coupling constant $g$, and we will verify that they are compatible with Eq.~(\ref{expw}).

The interacting parts of the SD Lagrangian is given by, 
\begin{equation}
\label{vert1}
L^{SD}_{int}\,=\,\frac{ig}{2}B^{\a}\left(
\phi\stackrel{\leftrightarrow}{D}_{\a}\bar{\phi}\right)-
\frac{g^2}{2}\bar{\phi}B^{\a}B_{\a}\phi\,,
\end{equation}

\noindent
while for the DMCS model, up to second order in $g$, we have two similar interaction terms,
\begin{eqnarray}
\label{vert2}
L^{DMCS}_{int}\,=\,\frac{ig}{4m}W^{\a}\left(\phi\stackrel{\leftrightarrow}
{D}_{\a}\bar{\phi}\right)
-\frac{1}{8}\frac{g^2}{m^2}\bar{\phi}\phi W^{\a}W_{\a}\,,
\end{eqnarray}

\noindent
together with the Thirring interaction,
\begin{equation}
\label{vert3}
L^{DMCS}_T\,=\,\frac{g^2}{64m^2}\int d^5 z (
\phi D^{\a}\bar{\phi}\phi D_{\a}\bar{\phi}
-2 D^{\a}\phi\bar{\phi}\phi D_{\a}\bar{\phi}+
 D^{\a}\phi \bar{\phi} D_{\a}\phi\bar{\phi})\,.
\end{equation}

\begin{figure}[htbp]
\includegraphics[width=0.5\textwidth]{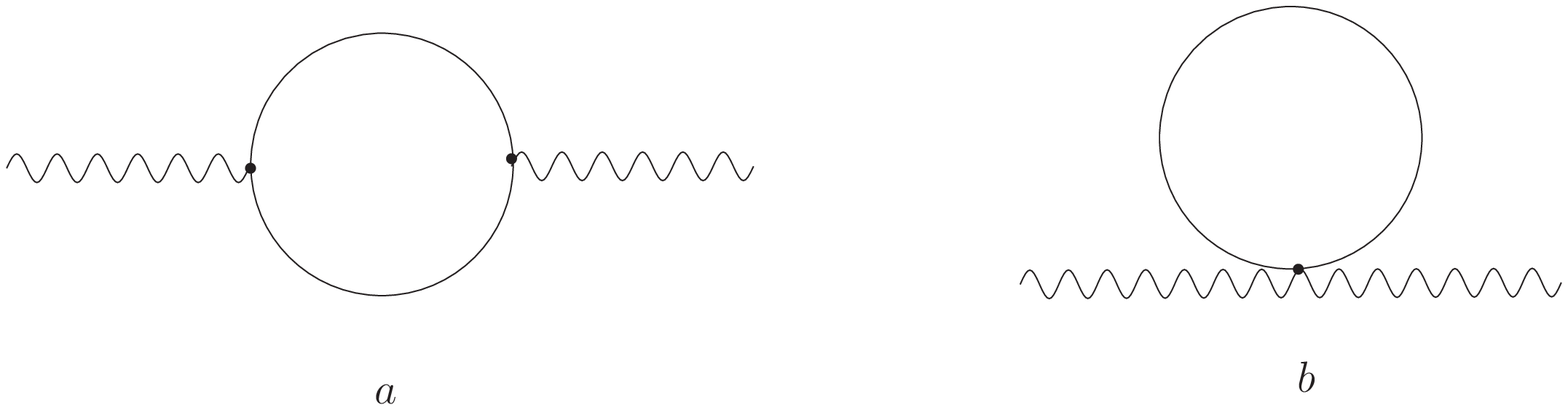}
\caption{\label{fig1}Contributions to the two-point function
of gauge field.}
\label{Fig1}
\end{figure}

We start by considering the first quantum corrections to the two-point function of the spinor superfields $B_\a$ and $W_\a$. In both cases, the relevant superdiagrams are those depicted in Fig.~\ref{fig1}, where each internal line stands for the $<\bar{\phi}\phi>$ propagator,
\begin{eqnarray}
<\bar{\phi}(k,\theta_1)\phi(k,\theta_2)>\,=\,-i\frac{D^2+M}{k^2+M^2}\delta_{12}\,,
\end{eqnarray}

\noindent
and the external wavy lines represent either the external $B^{\a}$ or $W^{\a}$ superfields. The evaluation of these diagrams yields a finite result, being equal to
\begin{align}
\label{stot}
iS^{SD}_1(p)\,=\,&-\frac{g^2}{4}
\int d^2\theta \int \frac{d^3k}{(2\pi)^3}
\,I(k,p)\,(k_{\gamma\beta}+MC_{\gamma\beta})
\nonumber\\&\times
\Big[(D^2B^{\gamma}(-p,\theta)) B^{\beta}(p,\theta)
+\frac{1}{2} D^{\gamma}D^{\alpha}B_{\alpha}(-p,\theta) B^{\beta}(p,\theta)
\Big],
\end{align}

\noindent
for the SD theory and, for the DMCS model,
\begin{align}
\label{stot2}
iS^{DMCS}_1(p)\,=\,&-\frac{g^2}{4}\frac{1}{4m^2}
\int d^2\theta \int \frac{d^3k}{(2\pi)^3}
\,I(k,p)\,(k_{\gamma\beta}+MC_{\gamma\beta})
\nonumber\\&\times
\Big[(D^2W^{\gamma}(-p,\theta)) W^{\beta}(p,\theta)
+\frac{1}{2} D^{\gamma}D^{\alpha}W_{\alpha}(-p,\theta) W^{\beta}(p,\theta)
\Big]\,,
\end{align}

\noindent
were we employed the notation $I(k,p)\,=\,\left\{(k^2+M^2)[(k+p)^2+M^2]\right\}^{-1}$. Notice that the contribution~(\ref{stot}) goes into~(\ref{stot2}) and vice-versa under the exchange of $B^{\a}$ by $\frac{W^{\a}}{2m}$. At the approximation we are working with, this is consistent with Eq.~(\ref{expw}), since the terms involving $\phi$ in the right hand side of~(\ref{expw}) contain additional powers of the coupling constant $g$. Hence, the duality between the SD and the DMCS models is maintained after the inclusion of the first quantum corrections induced by the diagrams in Fig.~\ref{fig1}.

\begin{figure}[htbp]
\includegraphics[width=0.7\textwidth]{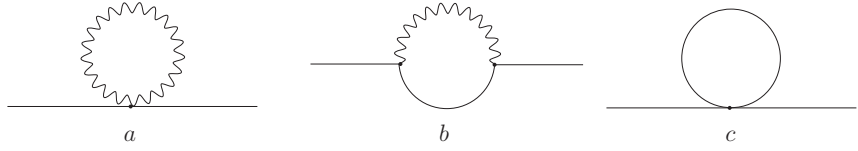}
\caption{\label{fig2}Contributions to the two-point function
of matter field.}
\label{Fig2}
\end{figure}

To further examine the persistence of the duality at the quantum level, we focus now on the corrections to the two-point function of the scalar superfield, which arise from the superdiagrams depicted in Fig.~\ref{fig2}. In the SD model, only the graphs \ref{fig2}a, \ref{fig2}b are present, whereas in the DMCS model the diagram \ref{fig2}c also contribute. The propagators of the superfield  strength $W^{\a}$ and the spinor superfield $B^{\a}$ are given  in Eqs.~(\ref{pw}) and~(\ref{pb}), respectively.

The evaluation of the supergraph in Fig.~\ref{fig2}a is the
simplest one. The result is the same both for the SD and for the DMCS theories, and it is given by
\begin{eqnarray}
\label{sc1}
iS_{2a}\,=\,\frac{g^2}{8m}\int\frac{d^3p}{(2\pi)^3}
\int d^2\theta\int\frac{d^3k}{(2\pi)^3}\,
\frac{1}{(k^2+4m^2)}\,\bar{\phi}(-p,\theta)\,\phi(p,\theta)
\end{eqnarray}

\noindent
(we note that the ``longitudinal'' term of the $B^{\a}$ propagator,
proportional to $\frac{1}{m^2p^2}$, do not contribute since it is proportional to $D^2D^{\a}D_{\a}\delta_{12}|_{\theta_1=\theta_2}=0$).

To calculate the  contribution from the graph in Fig.~\ref{fig2}b we regroup some terms in the propagator of the $B^{\a}$ superfield, which can be cast as
\begin{eqnarray}
\label{pb1}
<B^{\a}(-p,\q_1)B^{\b}(p,\q_2)>\,=\,\frac{i}{16}
\Big[\frac{(D^2+2m)D^{\b}D^{\a}}{m^2(p^2+4m^2)}-
2\frac{C^{\a\b}}{m^2}
\Big]\delta_{12}.
\end{eqnarray}

\noindent
The second term of this expression is constant, whereas the first term is equal to the propagator of $W^{\a}$ up to a factor $1/4m^2$. This difference is, however, compensated by the factor $1/4m^2$ in the quartic vertex of the DMCS model, Eq.~(\ref{vert2}), so that the contributions of the diagram~\ref{fig2}b in the DMCS model and the one corresponding to the first piece of the $B_\a$ propagator in the SD model are identical, and read
\begin{align}
\label{sc2}
iS_{2b}\,=\,&\frac{g^2}{64m^2}\int\frac{d^3p}{(2\pi)^3}
\int\frac{d^3k}{(2\pi)^3}\int d^2\q\frac{1}{(k^2+M^2)[(p-k)^2+4m^2]}
\nonumber\\
\times& \Big[-\bar{\phi}(-p,\theta) \left(D^2+2M+4m\right)\phi(p,\theta)(p-k)^2\nonumber\\
& + 2\bar{\phi}(-p,\theta)D^2\phi(p,\theta) \left(3kp-3k^2+4mM\right)\nonumber\\
&- 2\bar{\phi}(-p,\theta) \phi(p,\theta) \left(3Mp^2-mk^2-Mk^2-2Mpk-mpk\right)
\Big]\,.
\end{align}

\noindent
We stress that this expression is exact, including superficially divergent 
as well as finite parts.

The contribution from the second (constant) term of the $B_\a$ propagator in Eq.~(\ref{pb1}) can be found to be
\begin{equation}
\label{sc3}
S_{2b'}\,=\,\frac{g^2}{16m^2}\int\frac{d^3p}{(2\pi)^3}
\int d^2\theta\int\frac{d^3k}{(2\pi)^3}\,
\frac{1}{(k^2+M^2)}\,\bar{\phi}(-p,\theta)(D^2+M)\phi(p,\theta)\,.
\end{equation}

\noindent
There is no analog of such a constant term in the $<W^{\a}W^{\b}>$ propagator. However, in the DMCS theory there is the Thirring vertex, Eq.~(\ref{vert3}), which contributes to the two-point function of the $\phi$ superfield by means of the graph \ref{fig2}c. This contribution turns out to be exactly equal to $S_{2b'}$ in Eq. (\ref{sc3}). 

At the end of the day, we conclude that the first quantum corrections to the two-point vertex function of the scalar field for the self-dual
and the dualized Maxwell-Chern-Simons theories are identical, given by the sum of $S_{2a}$, $S_{2b}$ and $S_{2b'}$. This result confirms the duality between these two models when these quantum corrections are taken into account.

\section{Difficulties in a complete gauge embedding procedure}
\label{completege}

The gauge embedding procedure developed in Section~\ref{gaugeembedding} allowed us to obtain the theory defined by the action~(\ref{new}), dual to the SD model coupled to a scalar superfield, characterized by the ``restricted'' gauge symmetry $\delta A^{\alpha}=D^{\alpha}\epsilon$, where $\epsilon$ is an infinitesimal parameter and the matter superfield is kept untouched by the gauge transformation. A natural question is whether one can adapt this method to obtain a theory in which gauge transformations affect also the matter superfield, as it takes place in the usual supersymmetric electrodynamics~\cite{SGRS}. In this section, we develop the ``complete'' gauge embedding procedure, introducing 
Lagrange multipliers for both  the spinor and scalar superfields. 

From a formal viewpoint, the gauge embedding prescription is the following: starting with the Lagrangian $L(\Phi^i)$, where $\Phi^i$ is the set of the dynamical variables in the theory, then
\begin{equation} 
K_i\,=\,\frac{\delta L}{\delta\Phi_i} 
\end{equation} 

\noindent
are the corresponding Euler vectors. Next, if we denote the gauge transformation of each field as
$\Delta\Phi_i$, the total variation of the Lagrangian $L(\Phi^i)$ is
\begin{equation}
\delta L(\Phi^i)\,=\,K_i\Delta\Phi_i\,.
\end{equation}

\noindent
The first-order iterated Lagrangian is defined by
\begin{equation}
L^{(1)}(\Phi^i,\Lambda_i)\,=\,L-\Lambda_i K_i\,,
\end{equation}

\noindent 
where $\Lambda^i$ are the Lagrange multipliers, and the corresponding variation under a gauge transformation is
\begin{equation}
\delta L^{(1)}\,=\,K_i\Delta\Phi_i-K_i\delta\Lambda_i-\Lambda_i\delta
K_i\,.
\end{equation}

\noindent
To simplify this expression, we choose the Lagrange multiplier $\Lambda_i$ to change, under a gauge transformation, as $\delta\Lambda_i=\Delta\Phi_i$, and therefore 
\begin{equation}
\delta L^{(1)}\,=\,-\Lambda_i\delta K_i\,.
\end{equation}

\noindent
To cancel this variation we should augment $L^{(1)}$ by some
function of the Lagrange multiplier, $f(\Lambda)$, judiciously chosen so that the
second-order iterated Lagrangian,
\begin{equation}
\label{defl2}
L^{(2)}(\Phi^i,\Lambda_i)\,=\,L^{(1)}+f(\Lambda)\,,
\end{equation}

\noindent
is gauge invariant. The equation to be satisfied by $f(\Lambda)$ for this purpose is
\begin{equation}
\label{eqtwo}
 -\Lambda_i\delta K_i+f_{,i}(\Lambda)\Delta\Phi_i\,=\,0\,,
\end{equation}

\noindent
where $f_{,i}(\Lambda)= \frac{\partial f}{\partial\Lambda_i} $.

In summary, when Eq.~(\ref{eqtwo}) has a nontrivial solution $f(\Lambda)$, the gauge embedding method will provide us, in principle, with an invariant action given by~(\ref{defl2}). In Section~\ref{gaugeembedding}, we considered the situation in which the spinor superfield is transformed but not the scalar one, and in this case we were able to go through all steps of this procedure, obtaining the action~(\ref{new}). Now we turn to the case where the scalar superfield is also transformed and we will show that, even if we can find a nontrivial solution for~(\ref{eqtwo}), the application of the gauge embedding method turns out to be extremely cumbersome.

In the present case, besides the Euler vector for the spinor superfield $B^{\alpha}$, we introduce an Euler vector for the scalar superfield $\bar{\phi}$,
\begin{equation}
K\,=\,(D^2-M)\phi-igB^{\a}D_{\a}\phi-\frac{g^2}{2}B^{\a}B_{\a}\phi\,,
\end{equation}

\noindent
and the conjugated vector $\bar{K}$ for $\phi$. Correspondingly, we will
introduce the Lagrange multipliers $\Lambda$ and $\bar{\Lambda}$. The first-order iterated Lagrangian is given by
\begin{equation}
L^{(1)}\,=\,L-\Lambda^{\alpha}K_{\alpha}-\Lambda K-
\bar{\Lambda}\bar{K}\,,
\end{equation}

\noindent
and its variation under the infinitesimal gauge transformations 
\begin{eqnarray}
\label{var1}
\delta B_{\a} \,=\, D_{\a}\epsilon,\,
\delta\phi\,=\,ig\epsilon\phi,\,
\delta{\Lambda}\,=\,ig\epsilon{\phi}\,,
\end{eqnarray}

\noindent
is given by
\begin{eqnarray}
\delta L^{(1)}\,=\, -\Lambda^{\alpha}\delta
K_{\alpha}-\Lambda\delta K -\bar{\Lambda}\delta\bar{K}\,,
\end{eqnarray}

\noindent
after choosing the variations of the Lagrange multipliers as follows,
\begin{eqnarray}
\label{var}
\delta\Lambda_{\alpha}=D_{\alpha}\epsilon,\,
\delta\Lambda=ig\epsilon\phi,\,
\delta\bar{\Lambda}=-ig\epsilon\bar{\phi}\,.
\end{eqnarray}

\noindent
We also write the equation~(\ref{eqtwo}) in the case under consideration,
\begin{eqnarray}
\label{eqour}
-\Lambda^{\alpha}\delta K_{\alpha}-\L
\delta K -\bar{\L}\delta\bar{K}+
f_{,\alpha}(\Lambda)D_{\a}\epsilon+f_{\Lambda}
ig\epsilon\phi-f_{\bar{\Lambda}}ig\epsilon\bar{\phi}\,=\,0\,,
\end{eqnarray}

\noindent
where $f_{\Lambda}=\frac{\partial f}{\partial\Lambda},
f_{\bar{\Lambda}}= \frac{\partial f}{\partial\bar{\Lambda}},
f_{,\alpha}(\Lambda)= \frac{\partial
f}{\partial\Lambda^{\alpha}}$. 

Next, we evaluate the variations for the Euler vectors $K_{\alpha},K,\bar{K}$, starting with the spinor one,
\begin{eqnarray}
\label{dka}
\delta K_{\a}\,=\,8\m^2D_{\alpha}\epsilon+8(\delta\m^2)B_{\a}+\delta J_{\a}\,=\,(8\m^2+g^2\phi\bar{\phi})\delta\Lambda_{\a}\,=\, 8m^2D_{\a}\epsilon\,.
\end{eqnarray}

\noindent
It is interesting to compare this with Eq.~(\ref{compare}), to see the effect of the variation of the scalar superfield, which was absent in that case. As for the variation of the remaining Euler vectors, one finds
\begin{eqnarray}
\label{dk}
\delta K\,=\,ig(D^2-M) (\epsilon\phi) -i g D^{\alpha}\epsilon
D_{\alpha}\phi- g^2
B^{\alpha}(D_{\alpha}\epsilon)\phi-i\frac{g^3}{2}B^{\alpha}B_{\alpha}\epsilon\phi\,,
\end{eqnarray}

\noindent
and the complex conjugate of the above expression for $\delta \bar{K}$. 

Finally, inserting~(\ref{dk}) and~(\ref{dka}) into~(\ref{eqour}) and collecting the factors multiplying $D^2\epsilon$, $D_{\alpha}\epsilon$, and $\epsilon$, respectively, we found the condition~(\ref{eqour}) to be equivalent to the following set of equations,
\begin{subequations}
\begin{align}
\label{eq1}&\L\phi+\bar{\L}\bar{\phi} \,=\,0 \,, \\
&-8m^2\L_{\a}-2ig(\L D_{\a}\phi+\bar{\L}D_{\a}\bar{\phi}) -g^2B_{\a}(\L
\label{eq2}\phi+\bar{\L}\bar{\phi})+f_{,{\a}}(\L)\,=\,0 \,,\\ 
\label{eq3}&-ig[\L(D^2-M)\phi+\bar{\L}(D^2-M)\bar{\phi}]-\frac{ig^3}{2}B^{\a}B_{\a}
(\L\phi+\bar{\L}\bar{\phi})+f_{\L}ig\phi+f_{\bar{\L}}ig\bar{\phi}\,=\,0\,.
\end{align}
\end{subequations}

\noindent
Equation~(\ref{eq1}) is a constraint on the Lagrange multipliers $\Lambda,\bar{\Lambda}$, which can be inserted into Eqs.~(\ref{eq2}) and~(\ref{eq3}), to obtain 
\begin{subequations}
\label{eqq}
\begin{align}
& -8m^2\L_{\a}-2ig(\L
D_{\a}\phi+\bar{\L}D_{\a}\bar{\phi})+f_{,{\a}}(\L)\,=\,0 \,, \\
& -[\L D^2\phi+\bar{\L}D^2\bar{\phi}]+f_{\L}\phi+f_{\bar{\L}}\bar{\phi}\,=\,0\,.
\end{align}
\end{subequations}

\noindent
Equations~(\ref{eqq}) can actually be solved, and the solution reads
\begin{equation}
f(\Lambda)\,=\, -4m^2\L^{\a}\L_{\a}-4ig\L^{\a}(\L
D_{\a}\phi+\bar{\L}D_{\a}\bar{\phi})+\frac{\L^2
D^2\phi}{2\phi}+\frac{\bar{\L}^2 D^2\bar{\phi}}{2\bar{\phi}}\,,
\end{equation}

\noindent
which, going back to~(\ref{defl2}), gives the second-order Lagrangian 
\begin{eqnarray}
\label{thel2}
L^{(2)}&=&L-\Lambda^{\alpha}K_{\alpha}-\Lambda K-
\bar{\Lambda}\bar{K}-4m^2\L^{\a}\L_{\a}-4ig\L^{\a}(\L
D_{\a}\phi+\bar{\L}D_{\a}\bar{\phi})+\nonumber\\&+&\frac{\L^2
D^2\phi}{2\phi}+\frac{\bar{\L}^2 D^2\bar{\phi}}{2\bar{\phi}}\,.
\end{eqnarray}

\noindent
The corresponding equations of motion for $\L^{\alpha},\,\L,\,\bar{\L}$ are 
\begin{subequations}
\label{sys1}
\begin{align}
&-K^{\alpha}-8m^2\L^{\alpha}-4ig(\Lambda
D^{\alpha}\phi+\bar{\Lambda}D^{\alpha}\bar{\phi})\,=\,0 \,, \\
&-K-4ig\L^{\alpha}D_{\alpha}\phi-\frac{\L
D^2\phi}{\phi}\,=\,0 \,,\\
&-\bar{K}-4ig\L^{\alpha}D_{\alpha}\bar{\phi}-\frac{\bar{\L}
D^2\bar{\phi}}{\bar{\phi}}\,=\,0\,.
\end{align}
\end{subequations}

\noindent
Their solutions have the highly
cumbersome form
\begin{align}
\bar{\L}&\,=\,\Delta^{-1}\left[
\left(\bar{K}-ig\frac{K^{\b}D_{\b}\bar{\phi}}{2m^2}\right)
\left(\frac{2g^2}{m^2}D^{\alpha}\phi
D_{\alpha}\phi-\frac{D^2\phi}{\phi}\right)
-\left(K-ig\frac{K^{\b}D_{\b}\phi}{2m^2}\right)
\frac{2g^2}{m^2}D^{\alpha}\phi
D_{\alpha}\bar{\phi}\right]\,,\nonumber\\ 
\L&\,=\,\Delta^{-1}\left[
\left(K-ig\frac{K^{\b}D_{\b}\phi}{2m^2}\right)
\left(\frac{2g^2}{m^2}D^{\alpha}\bar{\phi}D_{\alpha}\bar{\phi}-\frac{D^2\bar{\phi}}{\bar{\phi}}\right)
-\left(\bar{K}-ig\frac{K^{\b}D_{\b}\bar{\phi}}{2m^2}\right)
\frac{2g^2}{m^2}D^{\alpha}\bar{\phi} D_{\alpha}{\phi}\right]\,,
\end{align}

\noindent
where
\begin{eqnarray}
\Delta\,=\,\left\vert\frac{2g^2}{m^2}D^{\alpha}\phi
D_{\alpha}\phi-\frac{D^2\phi}{\phi}\right\vert^2\,-\,
\frac{4g^2}{m^4}\left\vert D^{\alpha}\phi D_{\alpha}\bar{\phi} \right\vert ^2.
\end{eqnarray}

In principle, we could eliminate  $\L_\a$, $\L$ and $\bar{\L}$ from the second-order Lagrangian in Eq.~(\ref{thel2}) using their equations of motion but, as can be seen from the explicit solutions we have just quoted, in practice this would be extremely complicated. However, even without writing explicitly the effective Lagrangian obtained with our complete gauge embedding procedure, we note that, because of~(\ref{dka}), the Maxwell term in this effective Lagrangian would not appear with the field-dependent coefficient $1/\mu^2$. There is no simple way to relate the resulting effective theory to the dualized Maxwell-Chern-Simons we have obtained in Eq.~(\ref{new}). We see that, even if it does not provide us with a ``genuine'' gauge theory, the gauge embedding procedure adopted in Section~\ref{gaugeembedding} seems to be more adequate since it leads to a more tractable theory.

\section{Conclusions}
\label{conclusions}

We considered the dual equivalence between the supersymmetric self-dual and the supersymmetric Maxwell-Chern-Simons models in three spacetime dimensions. This duality was shown to take place for the free theories, leading to the known duality between the vector components of these superfields. To contemplate the situation where the interaction with matter is present, we used a gauge embedding method to build the dual to the supersymmetric SD model coupled to a scalar superfield, and found it to be a modified MCS theory, with an unusual field-dependent coupling for the Maxwell term, together with a Thirring interaction and a nonpolynomial ``magnetic'' coupling of the
matter to the gauge superfield. Then, we shown that the dual equivalence of these two models is maintained by the quantum corrections, at least in the one-loop approximation.

Also, we developed a prescription for a generalized gauge embedding procedure which allows one to obtain a dual theory invariant under gauge transformations for both gauge and scalar superfields. However, this result is at the most of academic interest, since it becomes very difficult to write explicitly the resulting effective Lagrangian in this case.

We close this paper by recalling recent discussions in the literature about whether the duality between self-dual and topological gauge theories is realized in noncommutative spacetimes~\cite{reviews,nccs,Harikumar:2005ry}, usually making use of the Seiberg-Witten (SW) map~\cite{SW}. In~\cite{Mariz:2003vx}, the question was analyzed without the recourse to the SW map, and it was argued that the noncommutative SD model is not dual to the noncommutative generalization of the MCS theory, but instead a modified noncommutative MCS dual model was unveiled (this conclusion is consistent with the analysis using the SW map in~\cite{Harikumar:2005ry}). One might hope that the gauge embedding method developed in this work could further elucidate these issues. However, after carefully applying the steps described in Section~\ref{gaugeembedding} in the noncommutative situation, one stops at the noncommutative version of Eq.~(\ref{24}), which assumes the form
\begin{equation}
\label{eqofmnc}
K_\a \, = \, 4 \left( \Lambda_\a * \m^2 + \m^{2} * \Lambda_\a \right)\,,
\end{equation}

\noindent
where $\m^2 = m^2 - \frac{g^2}{8}\,\phi * \bar{\phi}^2$, and the asterisk denotes the Groenewald-Moyal product. Therefore, in the noncommutative case, one cannot eliminate the Lagrange multiplier from the second order iterated Lagrangian using its equation of motion~(\ref{eqofmnc}). For the moment, this is a major stumbling block in applying the methods developed in this paper to the noncommutative version of the models studied here.

\vspace{1cm}

{\bf Acknowledgements.} A. Yu. P. is grateful to P. Minces for useful discussions. This work was partially supported by Funda\c{c}\~{a}o de Amparo \`{a} Pesquisa do Estado de S\~{a}o Paulo (FAPESP) and Conselho Nacional de Desenvolvimento Cient\'{\i}fico e Tecnol\'{o}gico (CNPq). The work of A. F. F. was supported by FAPESP,  project 04/13314-4. A. Yu. P. has been supported by CNPq-FAPESQ, DCR program (CNPq project 350400/2005-9).

\newpage

\end{document}